\documentclass[conference,a4paper]{IEEEtran}

\usepackage[mathletters]{ucs}
\usepackage[utf8x]{inputenc}

\usepackage{amsmath}
\usepackage{amsthm}
\usepackage{amssymb}
\usepackage{bussproofs}
\usepackage{stmaryrd}
\usepackage{comment}
\usepackage{color}
\usepackage{cite}
\usepackage{slashbox}
\usepackage{enumerate}
\usepackage{url}
\usepackage{listings}

\usepackage{xspace}
\usepackage{xstring}
\usepackage{ifthen}

\newtheorem{thm}{Theorem}

\newtheorem{dfn}{Definition}

\author{
\IEEEauthorblockN{Chantal Keller}
\IEEEauthorblockA{INRIA Saclay--Île-de-France at École Polytechnique\\
Email: Chantal.Keller@inria.fr}
\and
\IEEEauthorblockN{Marc Lasson}
\IEEEauthorblockA{ENS Lyon, Université de Lyon, LIP \\
UMR 5668 CNRS ENS Lyon UCBL INRIA\\
Email: marc.lasson@ens-lyon.org}
}

\title{The Refined Calculus of Inductive Construction: Parametricity and
Abstraction}

\DeclareUnicodeCharacter{10214}{\llbracket}
\DeclareUnicodeCharacter{10215}{\rrbracket}

\lstdefinelanguage{Coq}%
  {morekeywords={Variable,Inductive,CoInductive,Fixpoint,CoFixpoint,%
      Definition,Program, Lemma,Theorem,Axiom,Local,Save,Grammar,Syntax,Intro,%
      Trivial,Qed,Intros,Symmetry,Simpl,Rewrite,Apply,Elim,Assumption,%
      Left,Cut,Case,Auto,Unfold,Exact,Right,Hypothesis,Pattern,Destruct,%
      Constructor,Defined,Fix,Record,Proof,Induction,Hints,Exists,let,in,%
      Parameter,Split,Red,Reflexivity,Transitivity,if,then,else,Opaque,%
      Transparent,Inversion,Absurd,Generalize,Mutual,Cases,of,end,Analyze,%
      AutoRewrite,Functional,Scheme,params,Refine,using,Discriminate,Try,%
      Require,Load,Import,Scope,Set,Open,Section,End,match,with,Ltac,fun,forall,exists %
	},%
   sensitive, %
   morecomment=[n]{(*}{*)},%
   morestring=[d]",%
   literate={=>}{{$\Rightarrow$}}1 {>->}{{$\rightarrowtail$}}2{->}{{$\to\,\,\,\,\,$}}1
   {\/\\}{{$\wedge$}}1
   {|-}{{$\vdash$}}1
   {\\\/}{{$\vee$}}1
   {~}{{$\sim$}}1
   {'}{'}1
   {⟦}{{$\llbracket$}}1
   {⟧}{{$\rrbracket$}}1
  }[keywords,comments,strings]%

\DeclareMathOperator{\Prop}{\mathtt{Prop}}

\DeclareMathOperator{\Type}{\mathtt{Type}}
\DeclareMathOperator{\Set}{\mathtt{Set}}

\DeclareMathOperator{\Ind}{\mathtt{Ind}}

\DeclareMathOperator{\case}{\mathtt{case}}
\DeclareMathOperator{\fix}{\mathtt{fix}}

\DeclareMathOperator{\fingrp}{\mathtt{fingrp}}

\def\coq{\textsf{Coq}\xspace}

\def\cic{\textsf{CIC}\xspace}
\def\cicr{$\text{\textsf{CIC}}_{\text{\textsf{ref}}}$\xspace}

\newcommand{\rel}[3]{{⟦#1⟧}\,{#2}\,{#3}}

\newcommand{\arrlong}[1]{\overrightarrow{#1}}
\newcommand{\arrvar}[1]{\vec{#1}}

\newcommand{\we}{⊢_{*}}

\newcommand\arr[1]{
  \StrLen{$#1$}[\MyStrLen]
  \ifthenelse{\equal{\MyStrLen}{1}}
       {\arrvar{#1}\,\,\!}{\arrlong{#1}}}
\newcommand\arn[2]{
{\arrlong{#1}}^{#2}
}

\begin{document}

\lstset{breaklines=true, xleftmargin=0.3cm, xrightmargin=0.3cm,
  breakatwhitespace=true, mathescape=true, basicstyle=\ttfamily, %\scriptsize,
  numbers=none, frame=none, language = Coq}

\maketitle
\EnableBpAbbreviations

\begin{abstract}
  We present a refinement of the Calculus of Inductive Constructions in
  which one can easily define a notion of relational parametricity. It
  provides a new way to automate proofs in an interactive theorem prover
  like \coq.
\end{abstract}

\section{Introduction}

The Calculus of Inductive Constructions (\cic in short) extends the Calculus of
Constructions with inductively defined types.
% Si on est pas au courant de ça, alors on a de toute façon aucune chance de 
% pouvoir lire l'article :)
% It gives it a natural way of
%defining propositions and data types by induction, and facilitates extraction
%to ML-like languages. 
It is the underlying formal language of the \coq
interactive theorem prover~\cite{Coqdev11}.

In the original presentation, \cic had three kinds of sorts: the
impredicative sort of propositions $\Prop$, the impredicative sort of
basic informative types $\Set$, and the hierarchy of universes
$\Type_0$, $\Type_1$, \dots This presentation was not compatible with
the possibility to add axioms in the system, since it could lead to
inconsistencies~\cite{DBLP:conf/lics/Coquand86}. Nowadays, there is no
impredicative sort of basic informative types, and $\Set$ represents
$\Type_0$.

This does not fit well with one of the major original ideas about \cic:
the possibility to perform program extraction. Indeed, since the current
version of \cic does not separate informative types from non-informative
types, extraction needs to normalize its type to guess whether it should be 
erased or not, and this makes it very uneasy to
prove correct~\cite{DBLP:conf/cie/Letouzey08}.

In this paper, we propose a refinement of $\cic$ which reconciles
extraction with the possibility to add axioms to the system: \cicr, the
Refined Calculus of Inductive Constructions. The idea is to split the
$(\Type_i)_{i \in \mathbb{N}}$ hierarchy into two 
hierarchies $(\Set_i)_{i \in \mathbb{N}}$ and $(\Type_i)_{i \in
  \mathbb{N}^*}$, one for informative types and one for types without
computational content.

This calculus allows us to extend the presentation of
parametricity for Pure Types Systems introduced by Bernardy \emph{et
  al.}~\cite{DBLP:conf/icfp/BernardyJP10} to the Calculus of Inductive
Constructions. Parametricity is a concept introduced by
Reynolds~\cite{DBLP:conf/ifip/Reynolds83} to study the type abstraction
of system F, and the \emph{abstraction theorem} expresses the fact that
polymorphic programs map related arguments to related results. In \cicr,
we can define a notion of relational parametricity in which the
relations' codomains is the $\Prop$ sort of propositions.

\begin{comment} % Mes directeurs de these disent que c'est pas utile
After shortly presenting \cicr in Section~\ref{sec:cicr}, we define
parametricity in it and give the main result: the abstraction theorem
(Section~\ref{sec:param}). In Section~\ref{sec:application}, we give an
example of application of this result before concluding.
\end{comment}

%%% Oui... je le mets ici pour qu'il soit sur la dernière page !
\begin{figure*}
\begin{align*}
  Θ_I(\arn{Q}{p},T,\arn{F}{n}) = &\,λ\arn{(x:A)(x':A')(x_R:⟦A⟧\,x\,x')}{n}\,(a : I\,\arn{Q}{p}\,\arn{x}{n})
                 (a': I\,\arn{Q'}{p}\,\arn{x'}{n}) (a_R : ⟦I⟧\,\arn{Q\,Q'\,⟦Q⟧}{p}\,\arn{x\,x'\,x_R}{n} a\,a').\\
      & ⟦T⟧\,\arn{x\,x'\,x_R}{n}\,a\,a'\,a_R\,
                           \,(\case_I\,(a, \arn{Q}{p}, T, \arn{F}{n}))
                           \,(\case_{I}\,(a', \arn{Q'}{p}, T', \arn{F'}{n}))
\end{align*}
\caption{\label{fig}Relation parametricity for inductive types}
\end{figure*}

\section{\cicr: the Refined Calculus of Inductive
  Constructions}\label{sec:cicr}

The Refined Calculus of Inductive Constructions is a refinement of \cic where terms are generated by the same grammar as \cic:
\newcommand{\pouf}{\hspace{0.6em}}
$$
\begin{array}{c} A, B, P, Q, F \pouf := \pouf  x  
  \pouf|\pouf  s \pouf|\pouf ∀x:A.B  \pouf|\pouf λx:A.B \\
|\pouf  (A\,B) \pouf|\pouf  I \pouf|\pouf \case_I(A,\arr{Q}, P,
\arr{F})\pouf|\pouf c \pouf|\pouf \fix\,(x : A).B
\end{array} $$
where $s$ ranges over the set $\left\{\Prop\}∪\{\Set_i, \Type_{i+1} | i
  ∈ \mathbb{N} \right\}$ of \emph{sorts} and $x$ ranges over the set of
\emph{variables}. We write $\Ind^p(I:A, \arn{c:C}{k})$ to state
that $I$ is a well-formed inductive definition typed with $p$
parameters, of arity $A$, with $k$ constructors $c_1,\dots, c_k$ of
respective types $C_1,\dots,C_k$. 

\begin{comment} %% Est-ce vraiment utile de le préciser ? 
%% On verra si on s'en sert...
As usual, we will consider terms up to α-conversion and we denote by
$A[B/x]$ the term built by substituting the term $B$ to each free
occurrence of $x$ in $A$. The $β\iota$-reduction $\rhd_{β\iota}$ is
defined as in \cic, and we write $A≡_{β\iota}B$ to denote the
$β\iota$-conversion.
\end{comment}

A context $\Gamma$ is a list of pairs $x:A$ and the typing rules are the rules of \cic
(one can refer to~\cite{Coqdev11} for the complete set of rules), except to type sorts and 
dependent products. As for \cic, typing
fixpoints (for $\fix$) and elimination rules (for $\case$) is subject
to restrictions to ensure coherence. We present only the rules which 
are specific to our type system. Here are the three typing rules to type sorts: 
\newcommand{\pif}{\hspace{-0.4em}} % j'ai un peu honte :)
 \AXC{}
 \UIC{\pif$⊢ \Prop : \Type_1$\pif}

 \AXC{}
 \UIC{\pif$⊢ \Set_i : \Type_{i+1}$\pif}

 \AXC{}
 \UIC{\pif$⊢ \Type_i : \Type_{i+1}$\pif}
 \noLine
 \TIC{}\DP

The following three typing rules tell which products are authorized in 
the system. The level of the product is the maximum level of the 
domain and the codomain: 
\begin{center}
  \AXC{$Γ ⊢  A : r_i$}
  \AXC{$Γ, x : A ⊢  B : s_j$}
  \RightLabel{$(r,s) ∈ \{ \Type, \Set \}$}
  \BIC{$Γ ⊢ ∀x:A.B : s_{\max(i,j)}$} \DP
\end{center}

Quantifying over propositions does not rise the level of the product: 
\begin{center}
  \AXC{$Γ ⊢  A : \Prop$}
  \AXC{$Γ, h : A ⊢  B : s_i$}
  \RightLabel{$s ∈ \{ \Type, \Set \}$}
  \BIC{$Γ ⊢ ∀h:A.B : s_i$} \DP
\end{center}

And the sort $\Prop$ is impredicative, it means that products in $\Prop$
may be built by quantifying over objects whose types inhabit any sort:
\begin{center}
  \AXC{$Γ ⊢  A : s$}
  \AXC{$Γ, x : A ⊢  B : \Prop$}
  \RightLabel{$s ∈ \{ \Type, \Set, \Prop \}$}
  \BIC{$Γ ⊢ ∀x:A.B : \Prop$}
  \DP
\end{center}
\begin{comment}
The typing rules ensure the fact that the sort $\Type_i$ is populated only with
arities and higher-order functions that manipulate arities (an arity is a term
whose head normal form has the form $∀(x₁:A₁)\dots(x_n:A_n).s$ where $s$ is
either $\Prop$, $\Set_j$ or $\Type_j$ with $j < i$).
\end{comment}

Finally, as in \cic, the system comes with subtyping rules based on the following
inclusion of sorts (where $i < j$): 
 \AXC{}
 \noLine
 \UIC{$\Prop <: \Set_1$}

 \AXC{}
 \noLine
 \UIC{$\Set_i <: \Set_j$}

 \AXC{}
 \noLine
 \UIC{$\Type_i <: \Type_j$}

 \noLine
 \TIC{} \DP

One should note that \cicr easily embeds into \cic by mapping any $\Set_i$ and
$\Type_i$ onto the $\Type_i$ of \cic. The coherence of \cic thus implies the
coherence of \cicr.

\section{Parametricity}\label{sec:param}

We can define a notion of relational parametricity for \cicr.
\begin{dfn}[\label{Parametricity}Parametricity relation]
  For any inductive $\Ind^p(I:A, \arn{c : C}{k})$, we define a fresh
  inductive symbol $⟦I⟧$ and a family $(⟦c_i⟧)_{i=1...k}$ of fresh
  constructor names.

  The parametricity translation $⟦\bullet⟧$ is defined by induction on
  the structure of terms and contexts:
\begin{align*}
  ⟦\langle\rangle⟧ = &\,\langle \rangle \\
  ⟦Γ, x:A⟧ = &\,⟦Γ⟧, x:A, x':A',x_R :⟦A⟧\,x\,x' \\
  ⟦s⟧ = &\,λ(x:s)(x':s).x → x' → \hat{s} \\
  ⟦x⟧ = &\,x_R \\
  ⟦∀x\!:\! A.B⟧ = &\,λ(f:∀x:A.B)(f':∀x':A'.B').\\
  & ∀(x:A)(x':A')(x_R:\rel{A}{x}{x'}).\\
  &\rel{B}{(f\,x)}{(f'\,x')}\\
  ⟦λx:A.B⟧ = &\,λ(x:A)(x':A')(x_R:\rel{A}{x}{x'}).⟦B⟧ \\
  ⟦(A\,B)⟧ = &\,(⟦A⟧\,B\,B'\,⟦B⟧)\\
   ⟦\fix(x:A).B⟧ = & \left(\fix(x_R:⟦A⟧\,x\,x').⟦B⟧\right)\\
   &[\fix(x:A).B/x][\fix(x':A').B'/x']\\
   ⟦\case_I(M,\arn{Q}{p}, T,\arn{F}{n}&)⟧ = \case_{⟦I⟧}(⟦M⟧,\arn{Q, Q',
     ⟦Q⟧}{p},\\
   &Θ_I(\arn{Q}{p},T,\arn{F}{n}),\arn{⟦F⟧}{n})
\end{align*}
where $\hat{\Prop} = \hat{\Set_i} = \Prop$ and $\hat{\Type_i} = \Type_i$
and where $A'$ denotes the term $A$ in which we have replaced each variable
$x$ by a fresh variable $x'$. The definition of $\Theta_I$ is in \emph{Fig.~\ref{fig}}.

\end{dfn}

What is new with respect to previous works is the fact that relations over objects
of type $\Prop$ or $\Set_i$ have their codomain in $\Prop$ instead of 
higher universes. We also formally define parametricity for inductive
types.

Unfortunately, in order to prove the abstraction
theorem below, we need to restrict the strong elimination: we have to disallow the $\case$ destructions 
used to build objects whose types are of sort $\Type$ when the destructed inductive 
definition is not \emph{small} (\emph{small inductive definitions} are inductive 
definitions which constructors only have arguments of type $\Prop$ or $\Set$, 
see  \cite{springerlink:10.1007/BFb0037116}). 
We write $\we$ for the derivability  where strong elimination is
authorized only over small 
inductive definitions. % and forbidden otherwise.

\begin{thm}[\label{AbstractionInductive}Abstraction theorem] 
  If $Γ \we A : B$ then $⟦Γ⟧ \we A : B$, $⟦Γ⟧ \we A' : B'$, and
      $⟦Γ⟧ \we ⟦A⟧ : \rel{B}{A}{A'}$.
\end{thm}

\section{Applications}\label{sec:application}

A lot of so-called ``free theorems'' are consequences of the 
abstraction theorem and our framework is expressive enough to 
implement most examples that can be found in the literature % (e.g.~\cite{Wadler89, DBLP:conf/icfp/BernardyJP10}).
(see for instance \cite{Wadler89, DBLP:conf/icfp/BernardyJP10}). 

Here we propose a new example inspired by François Garillot's
thesis~\cite{Garillot11}, in which he remarks that polymorphic functions
operating on groups can only compose elements using the laws given by the
group's structure, and thus cannot create new elements. 

In our system, we may actually use parametricity theory to translate this
uniformity property. We take an arbitrary group structure
$\mathcal{H}$ defined by its carrier $\alpha : \Set_0$, a unit element, a composition law, an inverse
and the standard axioms stating that
$\mathcal{H}$ is a group. We define $\fingrp$ the type of all the
finite subgroups of $\mathcal{H}$ consisting of a list plus stability axioms.
Now consider any term $Z : \fingrp → \fingrp$ (examples of such terms abound:
e.g. the center, the normalizer, the derived subgroup\dots). The abstraction
theorem states that for any $R : \alpha → \alpha → \Prop$ compatible with the
laws of $\mathcal{H}$ and for any $G\,G' : \fingrp$, $⟦\fingrp⟧_R\,G \,G' →
⟦\fingrp⟧_R\,(Z\,G)\,(Z\,G')$ where $⟦\fingrp⟧_R$ is the relation on subgroups
induced by $R$. Given this, we can prove the following properties:
\begin{itemize} \item for any $G$, $Z\,G \subset G$ (if we take $R : x\,y
\mapsto x \in G$); \item for any $G$, for any $\phi$ a morphism of
$\mathcal{H}$, $\phi(Z\,G) = Z\,\phi(G)$ (if we take $R : x\,y \mapsto y =
\phi(x)$).  It entails that $Z\,G$ is a \emph{characteristic subgroup} of
$\mathcal{H}$.  \end{itemize}

For a complete \coq formalization of this, please refer to the 
online source code~\cite{implem12}.

\section{Conclusion}

The system presented here allows to distinguish clearly via typing which
expressions will be computationally meaningful after extraction. It allows us
to define a notion of parametricity for which relations lie in the sort of
propositions. We set here the theoretical foundation for an implementation of a
\coq tactic that constructs proof terms by parametricity. A first prototype of such
a tactic can be found online~\cite{implem12}.

\bibliographystyle{IEEEtran}
\bibliography{biblio-cheat}

\end{document}